\documentclass[10pt]{iopart}
\usepackage[]{psfig,amssymb}
\begin{document}
%\draft                       % this command makes pacs numbers print
%-------------------------------------------------------------------

\title{Structural properties of various sodium thiogermanate glasses through 
DFT-based molecular dynamics simulations}
\author{S\'ebastien Blaineau and Philippe Jund }
\address{
Laboratoire de Physicochimie de la Mati\`ere Condens\'ee, Universit\'e Montpellier 2, 
\\ Place E. Bataillon, Case 03, 34095 Montpellier, 
France
}
%\maketitle

\begin{abstract}
We present a study of the structural properties of (x)Na$_2$S-(1-x)GeS$_2$ glasses through DFT-based molecular dynamics 
simulations, at different sodium concentrations ($0<x<0.5$). We computed the radial pair correlation 
functions as well as the total and partial structure factors. We also analyzed the evolution of the corner- and 
edge-sharing intertetrahedral links with the sodium concentration and show that the sodium ions 
exclusively destroy the former. With the increase of the sodium concentration the ``standard'' FSDP 
disappears and a new pre-peak appears in the structure factor which can be traced back in the Na-Na partial 
structure factor. This self organization of the sodium ions is coherent with Na-rich zones that we find at 
high modifier concentration.
\end{abstract}
\pacs{PACS numbers: 61.43.Bn,61.43.Fs,71.15.Pd,63.50.+x }

%-------------------------------------------------------------------------

\section{Introduction}
Amorphous (x)Na$_2$S-(1-x)GeS$_2$ systems are interesting glasses,
presenting a high ionic conductivity at room temperature.
These materials can be used as model systems for efficient solid electrolytes 
even though practical applications are more often based on glasses 
containing lithium or silver ions for example.
Although the ionic conduction process has been clearly observed in sodium 
thiogermanate glasses, its microscopic origin is still not clearly understood.
Like in many other glassy systems, the atomic mechanisms responsible for 
these transport properties have not yet been clearly determined which explains 
the large number of studies dedicated to this topic in the past few years \cite{jund}.
In particular, it is interesting to determine if the modifier ions
follow preferential pathways inside the amorphous system \cite{greaves}, and if this 
process is dependent on the glass type. To that end, theoretical studies such as molecular 
dynamics (MD) simulations can provide detailed information concerning the microscopic properties 
of these chalcogenide glasses.\\
In previous studies we have analyzed the physical and chemical properties of GeS$_2$ glasses via 
Density Functionnal Theory (DFT)-based molecular dynamics simulations \cite{blaineau1,blaineau2,blaineau3}. 
The results obtained from our model were in very good agreement with the available experimental data. Subsequently we 
studied the vibrational and electronic properties of sodium thiogermanate (Na-Ge-S) glasses using the same 
description \cite{blaineau4}.
In the present work, we focus on the structural properties of (x)Na$_2$S-(1-x)GeS$_2$ amorphous systems 
using the same DFT-based model in order to determine the impact of the Na$^+$ ions on the structure of 
germanium disulfide glasses. It should be noted that there is a lack of experimental results in the
literature concerning these doped chalcogenide glasses. Thus no extensive comparison with experiments is possible 
at this time, and our results should be considered as a first step in the study of the structural 
properties of (x)Na$_2$S-(1-x)GeS$_2$ glasses. \\
This paper is organized as follows: section II is dedicated to the description of the theoretical foundations 
of our model. Subsequently, we present the results obtained in our simulations in section III. 
Finally, section IV contains the major conclusions of our work.

\section{Model}
The mathematical model we have used was developed by Sankey and Niklewski \cite{sankey}. It is included in a 
first-principles type molecular-dynamics code, called {\sc{fireball96}}, that is based on the 
DFT \cite{hohenberg-kohn}, within 
the Local Density Approximation (LDA) \cite{ceperley,perdew}. 
A tight-binding-like linear combination of pseudoatomic orbitals, satisfying the atomic self-consistent 
Hohenberg-Kohn-Sham equations \cite{kohn-sham}, is used to determine the electronic eigenstates of the system. 
A basis set of one $s$ and three $p$ pseudo-orbitals, slightly excited since they vanish outside a cut-off radius of 
$5a_o$ (2.645 \AA), is required.\\
The pseudopotential approximation is used to replace the core electrons with an effective potential which acts 
on the valence electrons, and Hamman-Schluter-Chiang pseudopotentials were used in this work \cite{schluter}. 
The Hamiltonian 
is calculated through the use of the Harris functional \cite{harris}, which avoids long self-consistent calculations by 
a zero$^{th}$ order approximation of the self-consistent density. Further details on the method can be found in 
the original paper \cite{fireball}. 
 Periodic boundary conditions are applied in order to
limit surface effects, and only the $\Gamma$ point is used to sample the
Brillouin zone. This model has given excellent results in several different
chalcogenide systems over the past few years \cite{blaineau1,drabold,junli}.\\
In the present work we computationally melted a crystalline $\alpha$-GeS$_2$ configuration 
containing 258 particles at 2000K during 60 ps (24000 timesteps) until 
we obtained an equilibrated liquid. Subsequently we randomly replaced GeS$_4$ tetrahedral units by 
Na$_2$S$_3$ units, in order to obtain a given concentration of modifier without affecting the global 
electrical neutrality of the system. This procedure has been used in a similar manner in SiO$_2$ glasses \cite{jund1}, 
and keeps the total number of atoms, N, unchanged (258 particles in this work).\\
Following this scheme, we generated eight (x)Na$_2$S-(1-x)GeS$_2$ samples with different sodium concentrations (x= 
0, 0.015, 0.03, 0.06, 0.11, 0.2, 0.33 and 0.5). The bounding box was rescaled for each sample so that 
the density matched its experimental counterpart (with cell lengths ranging from 19.21 \AA~for  x=0\cite{boolchand} to 18.3 
\AA~for x=0.5\cite{ribes}). Since the experimental densities are not exactly the equilibrium densities of the
model, the pressure inside the samples is not exactly equal to zero. Nevertheless it is extremely small
($<$ 0.4 GPa) demonstrating the quality of the model. The resulting system 
was subsequently melted at 2000K during 60 ps in order for the system to lose memory of the previous artificial 
configuration. When the liquid reached an equilibrium state, the systems were quenched at a quenching rate of 
6.8$\times$10$^{14}$~K/s, decreasing the temperature to 300K, passing through the glass transition temperature 
T$_g$. The samples were then relaxed at 300K over a period of 100 ps. In order to improve the statistics of our results, 
five samples, starting from independent liquid configurations, were generated for each concentration. 
The results presented here have been averaged over these five samples.

\section{Results}

In a previous work \cite{blaineau3}, we determined that the atomic 
charges in GeS$_2$ glasses using a L\"owdin description \cite{lowdin} are equal to +0.94 for Ge atoms and 
-0.47 for the S particles. Even though homopolar bonds exist in this glass \cite{blaineau2}, no Ge-Na bonds 
were seen in our samples at the concentrations studied here ($0< x \le 0.5$). 
Therefore we can study the impact of the sodium ions on the 
short-range order of germanium disulfide glasses by analysis of the S-Na bonds exclusively.\\
Among the radial pair correlation functions $g_{\alpha \beta}(r)$ defined as:
\begin{equation}
g_{\alpha \beta}(r)=\frac{V}{4 \pi r^{2} N_{\alpha} dr}~dn_{\beta}
\end{equation}
for the different ($\alpha$,$\beta$) pairs, we can hence concentrate on the $g_{SNa}(r)$ function.
It can be seen in Fig.1 that the $g_{SNa}(r)$ for S-Na pairs does not change 
significantly with the concentration of the sodium modifier. The first-neighbor peak 
appears at 2.4 \AA~for all concentrations ($0< x\le 0.5$), which is slightly inferior
to experimental data concerning crystalline sodium thiogermanate systems 
(2.71 \AA~\cite{foix}). However we tested several glassy samples with
the program {\sc{siesta}} \cite{siesta} in a self-consistent $ab~initio$ 
description, using the most accurate basis set available in the code for the description of
the atomic orbitals (Double Zeta Polarized + Generalized Gradient 
Approximation): the results were very similar (d$_{S-Na}=2.45$ \AA) to those obtained with 
{\sc{fireball96}}.\\
The different coordination numbers $n_{\alpha \beta}$ can also be calculated. They correspond to the 
average number of $\beta$ neighbors of a given $\alpha$ atom within a sphere of radius r$_{min}$, 
r$_{min}$ being the first minimum of the corresponding $g_{\alpha \beta}(r)$.
We show in Fig.2 the evolution of the coordination number of the sulfur atoms 
with the three elements present in our system. It can be seen that the value of $n_{S-Ge}$ 
decreases with the addition of sodium modifier from 1.95 for x=0 to 1.3 for x=0.5. 
This indicates that the Na$^+$ ions break several inter-tetrahedral connections 
in order to be connected to the sulfur atoms, which explains the increasing value of $n_{S-Na}$ (2.45 for x=0.5).
The value of $n_{S-S}$ relative to homopolar bonds remains constant without any dependency with the sodium
concentration. Finally the average coordination number of the sodium atoms relative to the sulfur 
atoms, $n_{Na-S}$, has been found close to 3.27 for x=0.5. In fact, the Na$^+$ particles in the system are 
connected to 3 or 4 sulfur atoms, which is consistent with the results obtained by Foix $et~al.$ in 
cluster simulations of Na$_2$GeS$_3$ glasses using a Hartree-Fock model. \cite{foix}.\\
We reported in a previous study \cite{blaineau2} that 14.53$\%$ of the sulfur atoms in 
glassy GeS$_2$ systems are non-bridging (i.e. connected to only one Ge atom). These bond defects have
been found \cite{blaineau3} to create a negatively charged electronic environment around the
terminal sulfur. Therefore it is interesting to study the evolution of these non-bridging
sulfur particles with the introduction of Na$^+$ cations. We distinguish here between the 
terminal sulfur atoms connected to (a) zero, (b) one, (c) two or (d) three sodium ions, and represent in 
Fig.3  the evolution of these structural entities in (x)Na$_2$S-(1-x)GeS$_2$ for $0 \le x \le 
0.5$. \\
One can see that the proportion of terminal sulfur atoms that are not linked to a sodium ion decreases 
significantly as soon as x is non zero. This variation is counterbalanced by the increase of the 
proportion of terminal sulfur atoms connected to one sodium atom, which reaches approximately the same 
value of $\approx$15$\%$ of the S particles in the sample. We can therefore deduce that, as expected, 
the non-bridging sulfurs 
present in $g$-GeS$_2$ significantly attract the Na$^+$ ions. Between $x=0.11$ and $x=0.33$ these structural 
entities decrease (Fig.3(b)), and are counterbalanced by the increase of terminal sulfur connected to 
two Na ions (Fig.3(c)), that also reaches the value of $\approx$15 $\%$. Finally, for $x > 0.33$ the proportion of 
terminal sulfurs connected to three sodium ions reaches $\approx$15 $\%$ as in the aforementioned cases. 
This shows that the non-bridging 
sulfur atoms observed in $g$-GeS$_2$ attract $several$ Na$^+$ ions in sodium thiogermanate glasses.\\
The study of the short-range charge deviation in $g$-GeS$_2$ has also revealed the existence of ``positively'' 
charged zones, which are mainly caused by 3-fold coordinated sulfur and homopolar bonds~\cite{blaineau3}.
We find that these positively charged zones remain basically unchanged in (x)Na$_2$S-(1-x)GeS$_2$ glasses, and that 
the Na$^+$ ions never connect to 3-fold coordinated sulfur atoms. Hence negatively and positively charged zones in 
$g$-GeS$_2$ appear to attract and repel the Na$^+$ cations respectively. This observation, which seems to be 
self-evident, is in fact relevant since all the different samples have been melted, quenched, and relaxed 
{\em independently}. Despite that, we find similarly charged zones in comparable proportions in all 
of these samples.\\
The decrease of the $n_{S-Ge}$ coordination number with increasing sodium content observed in Fig.~2 shows that 
an increasing number of  intertetrahedral bonds are destroyed by the addition of sodium ions. 
Since two different types of intertetrahedral links exist in glassy GeS$_2$ (edge-sharing and corner-sharing links), 
it is interesting to determine which type of S-Ge connections are destroyed by the introduction of the Na$^+$ ions.
These two types of intertetrahedral connections can be distinguished in the radial pair correlation function of
germanium-germanium pairs \cite{blaineau2}, leading to Ge-Ge distances equal to 2.91 \AA~(edge-sharing) and 
3.41 \AA~(corner-sharing). In Fig.~4 we present the {\em total} number of corner-sharing and edge-sharing links 
in the different samples as a function of sodium concentration. It can clearly be seen that exclusively  
corner-sharing links are destroyed by the introduction of sodium cations whereas the number of 
edge-sharing links remains constant for all sodium concentrations. This result in not predictable 
{\em a priori} since very different results may be obtained depending on the glass and the modifier.  
For example, experimental studies on chalcogenide SiS$_2$ glasses have shown that Li$^+$ 
cations destroy exclusively edge-sharing links \cite{micoulaut,pradel}, while the addition of
Na$^+$ ions was found to break corner-sharing and edge-sharing links in equal proportions \cite{pradel2}. Here we find that 
sodium ions destroy exclusively corner-sharing links in GeS$_2$ glasses but this result still has to 
be confirmed (or refuted) experimentally.\\
In order to analyze the global impact of the sodium ions on the structure of thiogermanate glasses, 
we studied next the total static structure factor $S(q)$ of our (x)Na$_2$S-(1-x)GeS$_2$ glasses.
The total static structure factor was computed using the following formulae:
\begin{equation}
S(q) = 1 + 4 \pi \frac{N}{V} \int_0^{r_{max}} ( g(r) - 1 ) \frac{\sin(qr)}{qr} r^2 dr 
\end{equation}
where
\begin{equation}
g(r) = \frac{ \left[\displaystyle \sum_{\alpha,\beta} c_{\alpha} b_{\alpha} c_{\beta} b_{\beta} g_{\alpha \beta}(r) \right] }
{\left[ \displaystyle  \sum_{\alpha} c_{\alpha} b_{\alpha} \right]^2 }
\end{equation}
In the above equations, $r_{max}=L/2$ where $L$ is the edge of the box, the $g_{\alpha \beta}(r)$ are the radial pair correlation functions ($\alpha,\beta=$ Ge,S,Na), $c_{\alpha}$ and $c_{\beta}$  are the concentrations of the species $\alpha$ and $\beta$ and $b_{\alpha}$ and $b_{\beta}$ are their scattering lengths taken equal to 8.18 fm for Ge, 2.84 fm for S and
to 3.63 fm for Na \cite{Sq}.
The results for $S(q)$, calculated using the above definition, are shown in Fig.5 for the different values of $x$ (even 
though these results are directly comparable to neutron scattering experiments, no experimental data is available to our 
knowledge).\\
The first sharp diffraction peak (FSDP), which is a signature of amorphous 
materials, appears at $q=$1 \AA$^{-1}$ in $g$-GeS$_2$ \cite{blaineau1}. This peak reveals the 
existence of a structural order on a length scale $d$ corresponding to $d=2\pi /q\approx 
6.3$~\AA~in germanium disulfide glasses. In sodium thiogermanate glasses, it can be seen that the 
intensity of the FSDP decreases significantly with $x $ for $x\leq 0.2$. This decrease of the
FSDP  has also  been observed experimentally in silver thiogermanate glasses \cite{pradel}. 
This means that the introduction of  ions disturbs the intermediate range order of the 
glassy matrix by breaking corner-sharing intertetrahedral connections. 
However for $x\geq 0.2$, a new feature appears in the structure factor at 
a lower value of $q$ ($q=0.5$~\AA$^{-1}$). For x=0.5 this pre-peak reaches almost the same intensity
than the original FSDP, which almost completely vanishes and shifts to a slightly higher $q$ value. In order to
find the origin of this new peak we calculated the partial structure factors (computed from the radial pair 
correlation functions $g_{\alpha \beta}(r)$) in the different samples, 
and in particular the Na-Na partial structure factor (Fig.6). We find that the sodium-sodium pairs are 
mainly responsible for the pre-peak observed in Fig.5. It can indeed be seen that a peak emerges 
at $q=0.5$~\AA$^{-1}$ for $x \geq 0.11$
and subsequently grows with increasing $x$. 
To analyze the structural arrangement of the sodium atoms, which are obviously not distributed
homogeneously inside the simulation box, we present the density of sodium atoms in a 
(0.33)Na$_2$S-(0.66)GeS$_2$ averaged over the relaxation time in Fig.7.
To that end, we divide the bounding box into 1000 (10$\times$10$\times$10) elementary volume elements. 
We subsequently calculated the number of Na atoms located in each specific volume element, and integrated this
quantity over the relaxation time. Hence each sphere represented in Fig.7 is located
at the center of a volume element, and its size increases (becomes more red in color) with the local 
Na density (the largest sphere and thus the highest density corresponds to 0.154 \AA$^{-3}$).
This figure shows that our sodium thiogermanate sample contains several zones which have a 
high concentration of Na$^+$ and other zones which have a poor concentration of Na$^+$ 
ions. When we analyzed the local structure around the sodium atoms we found that most
of them are in a Na$_2$S-like environment, since for $x$=0.5, 23.5 $\%$ of all the 
sulfur atoms are {\em exclusively} connected to Na ions,  i.e. without any germanium atom 
as their nearest-neighbors. These sodium rich zones could be the static ``remainders'' of conduction 
channels that could be observed at higher temperatures similarly to what has been reported         
in sodium enriched silicate glasses via {\em classical} molecular dynamics simulations and inelastic 
neutron scattering experiments \cite{jund1,meyer}. If we accept this hypothesis, then, since the new pre-peak 
appears only for high sodium concentrations (see Fig.6), conduction channels would be absent
at low sodium content and this could explain the experimentally observed increase in 
the sodium diffusion constant when the sodium concentration is greater than $x$=0.10 \cite{annie}. Unfortunately, 
the DFT-based model that we used to perform these simulations are too computationally expensive making 
 it difficult to reach the simulation time required to study the diffusion of the Na$^+$ ions, 
and to analyze the evolution of these zones from a dynamical point of view. This is the reason why 
the sodium rich zones presented in Fig.7 cannot be directly assimilated to the diffusion channels.\\
Most of theses results have no experimental counterparts and therefore need still to be confirmed or invalidated,
 but neutron diffraction studies on these systems have been performed and the results should be available in a not so 
distant future.\\
To pursue this study further it would also be interesting to analyze the influence of the nature of the modifier 
on the results. Thus similar studies on silver thiogermanate glasses are currently in progress. 

\section{Conclusion}
We have studied the structural properties of sodium thiogermanate glasses via DFT-based molecular dynamics 
simulations for eight different modifier concentrations. We find that the structure of these chalcogenide 
glasses is significantly influenced by the extended positively and negatively charged zones observed in 
amorphous GeS$_2$ \cite{blaineau3}. As expected, the sodium cations are mainly attracted by the negative zones and avoid 
the positive regions. At high sodium concentrations, the Na$^+$ ions therefore become localized in space, 
and zones containing high and low density of sodium cations can be observed. These Na$_2$S-rich zones could be 
the low temperature analogs of the ``conduction channels'' observed in silicate glasses, where the diffusing ions follow 
preferential pathways. We have also determined that the addition of sodium cations inside the (x)Na$_2$S-(1-x)GeS$_2$ 
system destroys exclusively the corner-sharing links for $x\le 0.5$.\\
The structure factor of the different samples shows that the FSDP that appears in  amorphous GeS$_2$ at 
$q=$1 \AA$^{-1}$, decreases significantly with the addition of sodium ions. At high sodium concentrations ($x \ge 0.11$), 
a new pre-peak appears at $q=0.5 \AA^{-1}$, which corresponds to a distance in real space of 12.6~\AA. 
The Na-Na partial $S(q)$ shows that the Na-Na pairs are responsible of this pre-peak indicating a non
homogeneous distribution of the Na$^+$ ions inside the glass which is consistent with the existence 
of the aforementioned sodium rich zones.\\
\newpage
\hspace*{-0.65cm}{\bf Acknowledgments:} Parts of the simulations have been performed on the 
computers of the ``Centre Informatique National de l'Enseignement Sup\'erieur'' (CINES) 
in Montpellier.\\ Dr. N.A. Ramsahye is acknowledged for proof reading the manuscript.\\
\hrule

\newpage
\vspace{-3cm}
\begin{figure}
\centerline{\psfig{file=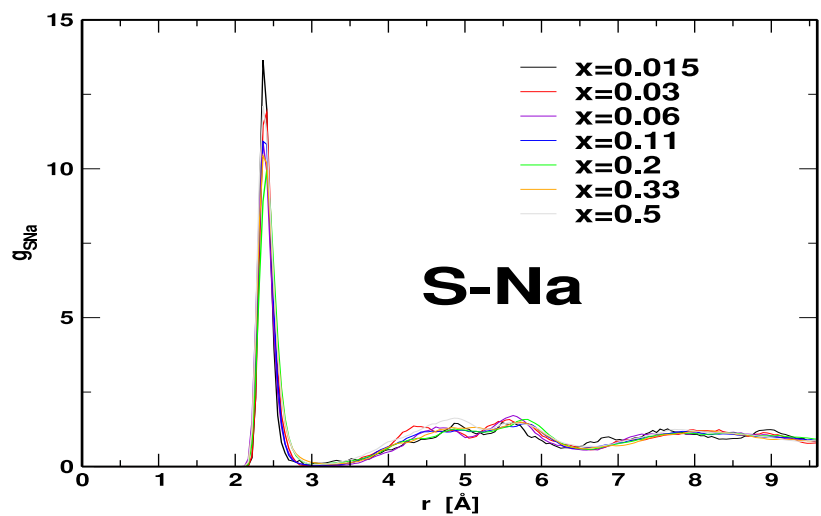,width=12cm,height=11cm}}
\caption{Radial S-Na pair distribution functions of (x)Na$_2$S-(1-x)GeS$_2$
glasses for $0<x\le 0.5$}
\label{fig1}
\end{figure}
\vspace{-3cm}
\begin{figure}
\centerline{\psfig{file=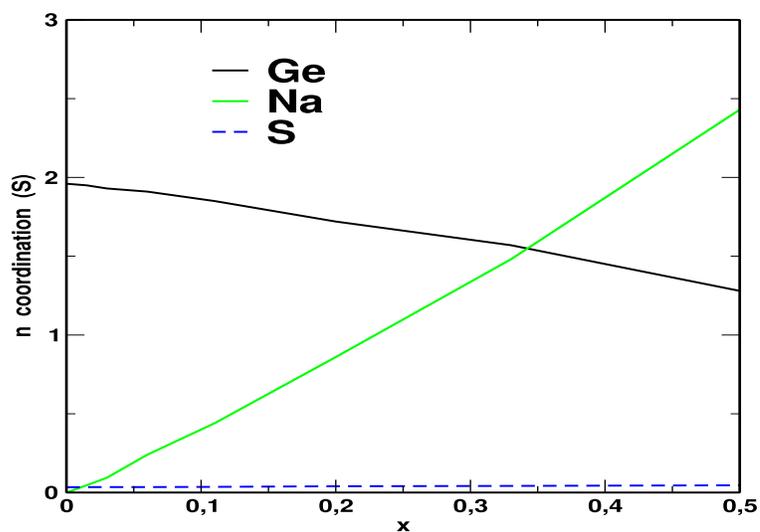,width=12cm,height=11cm}}
\caption{Coordination number of sulfur atoms in sodium thiogermanate samples, plotted
against x, the concentration of sodium atoms.}
\label{fig2}
\end{figure}

\newpage

\begin{figure}
\centerline{\psfig{file=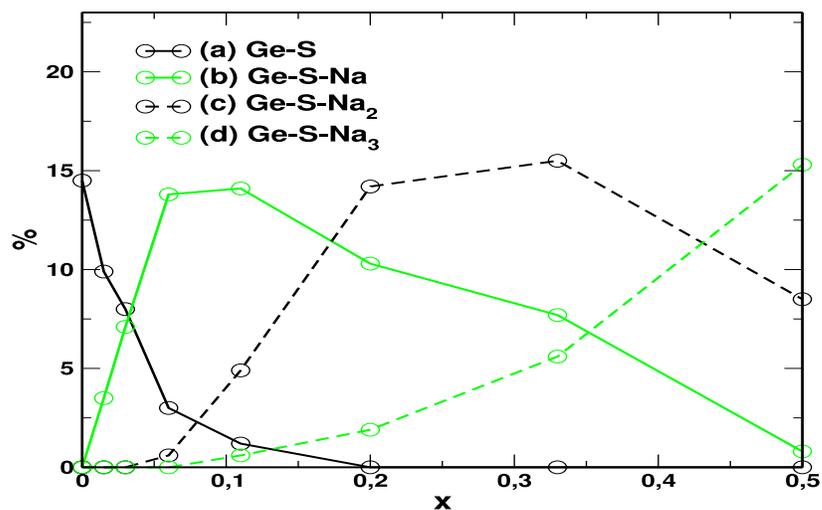,width=12cm,height=11cm}}
\caption{Proportion of non-bridging sulfur atoms in (x)Na$_2$S-(1-x)GeS$_2$ glasses. We distinguish the S atoms connected to (a) zero, (b) one, (c) two and (c) three Na$^+$ ions}
\label{fig3}
\end{figure}

\begin{figure}
\centerline{\psfig{file=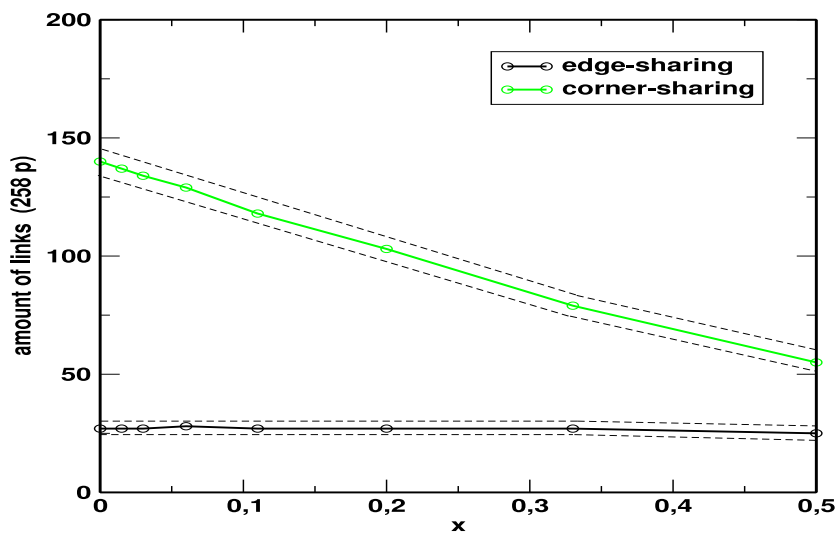,width=13cm,height=11cm}}
\caption{Number of edge-sharing and corner-sharing links in (x)Na$_2$S-(1-x)GeS$_2$ glasses, plotted 
against x, the concentration of sodium atoms.}
\label{fig4}
\end{figure}

\newpage

\begin{figure}
\centerline{\psfig{file=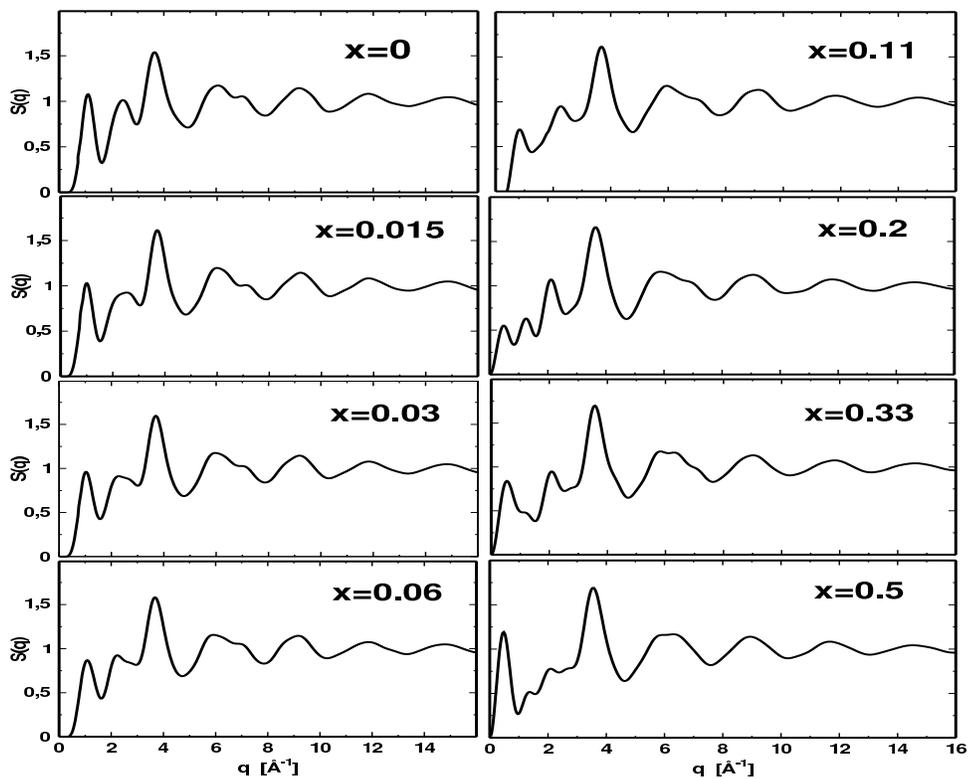,width=13cm,height=11cm}}
\caption{The simulated static structure factors of the different sodium thiogermanate glasses.}
\label{fig5}
\end{figure}

\begin{figure}
\vspace*{-1.5cm}
\centerline{\psfig{file=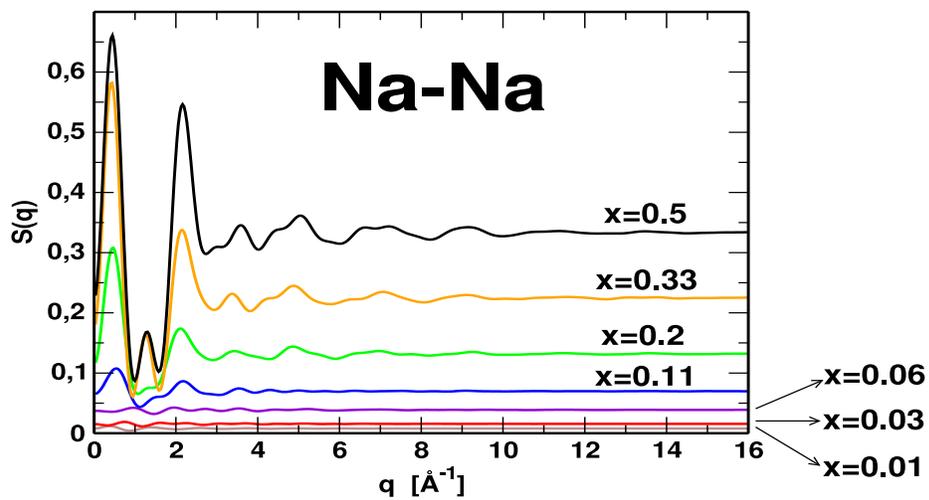,width=13cm,height=14cm}}
\vspace*{-2cm}
\caption{Partial S(q) for Na-Na pairs at different sodium concentrations versus q.}
\label{fig6}
\end{figure}

\newpage

\begin{figure}
\centerline{\psfig{file=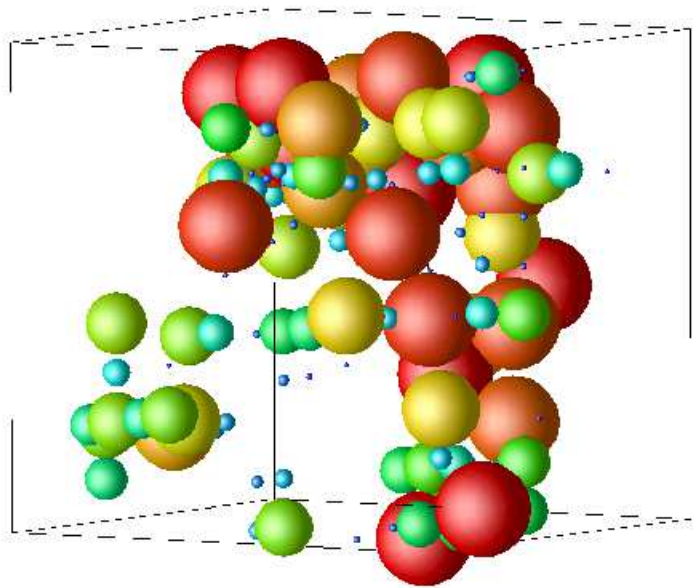,width=13cm,height=13cm}}
\vspace*{-2cm}
\caption{Illustration of the local Na density in a (0.33)Na$_2$-(0.66)GeS$_2$ sample integrated over the relaxation time (100 ps). The largest sphere corresponds to a Na density of 0.154~\AA$^{-3}$.}
\label{fig7}
\end{figure}

\end{document}